\documentclass[twocolumn,pre,aps,showpacs,amsmath,amssymb]{revtex4-1}
\usepackage{graphicx}
\usepackage{bm}
\usepackage{epstopdf}
\begin{document}

\title{Exactly solvable nonequilibrium Langevin relaxation of a trapped nanoparticle}
\author{Domingos S. P. Salazar$^{1}$}
\email[]{salazar.domingos@gmail.com}
\author{S\'ergio A. Lira$^{2}$}
\email[]{sergio@fis.ufal.br}
\affiliation{$^1$ Unidade Acad\^{e}mica de Educac\~{a}o a Dist\^{a}ncia e Tecnologia, Universidade Federal Rural de Pernambuco, Recife, Pernambuco 52171-900 Brazil \\
$^2$ Instituto de F\'{\i}sica, Universidade Federal de Alagoas, Macei\'o, Alagoas 57072-900 Brazil}


\begin{abstract}
In this work, we study the nonequilibrium statistical properties of the relaxation dynamics of a nanoparticle trapped in a harmonic potential. We report an exact time-dependent analytical solution to the Langevin dynamics that arises from the stochastic differential equation of our system's energy in the underdamped regime. By utilizing this stochastic thermodynamics approach, we are able to completely describe the heat exchange process between the nanoparticle and the surrounding environment. As an important consequence of our results, we observe the validity of the heat exchange fluctuation theorem (XFT) in our setup, which holds for systems arbitrarily far from equilibrium conditions. By extending our results for the case of $N$ noninterating nanoparticles, we perform analytical asymptotic limits and direct numerical simulations that corroborate our analytical predictions.
\end{abstract}
\maketitle

\section{Introduction}
\label{intro}

In the last two decades, the academic interest in small systems arbitrarily far from equilibrium conditions have increased considerably~\cite{Bustamante2005,Seifert2012}.
This can be associated to the enormous development in the fields of nanotechnology and molecular biology, which deal with physical systems of a scale in which thermal
fluctuations play a major role in their dynamics and statistics~\cite{Velasco2011}. As a matter of fact, at the nanoscale level, microsystems do not behave as a rescaled
version of their macroscopic counterpart and, in order to proper describe them, new universal laws must be assumed instead of the usual thermodynamics approach~\cite{Seifert2016}.

In this scenario, fluctuation theorems (FTs) have been theoretically derived~\cite{Evans1993,Jar1997,Crooks1999} and studied \cite{Crooks2007,Evans2015} in order to
state probability ratios between entropy generating processes and corresponding entropy consuming trajectories at nonequilibrium. These exact relations go beyond linear
response theory, thus reaching the regime of systems arbitrarily far from equilibrium. Moreover, FTs have been successfully verified in
experiments~\cite{Evans2002,Carberry2004,Ritort2005,Gieseler2012,Blickle2011} and computer simulations~\cite{Sivak2013}. Among them, in this work we highlight the importance of the
exchange fluctuation theorem (XFT)~\cite{Jar2004} applied to the heat exchanged between two systems $1$ and $2$
\begin{equation}
\label{FT1}
\ln{\frac{P_{t}(+Q)}{P_{t}(-Q)}} = \Delta \beta Q,
\end{equation}
where $P_{t}(+Q)$ denotes the probability that a net heat $Q$ to be transferred from system $1$ to system $2$ during a specified time interval $t$, while $P_{t}(-Q)$
represents the probability of $Q$ to flow from $2$ to $1$. The parameter $\Delta \beta = (k_B T_{2})^{-1} - (k_B T_{1})^{-1}$ is the difference between the inverse
temperatures at which the systems are prepared and $k_{B}$ is the Boltzmann constant. We point out that Eq.~(\ref{FT1}) consists on a general statistical statement about
heat exchange between classic or quantum systems and, contrary to other FTs, it is independent of the specific definition of the system's entropy.

Despite the increasing interest in nonequilibrium stochastic thermodynamics, few exactly solvable models of systems arbitrarily far from equilibrium are available in the literature, except for the case where nonequilibrium steady states are reached~\cite{Seifert2012,Gieseler2012}. In order to fill this gap, in this work we use a statistical thermodynamics analysis to obtain the analytic solution to a simple paradigmatic nanosystem: a Brownian nanoparticle optically trapped in a harmonic potential and immersed in a thermal bath. Hereafter, we investigate this system via the Langevin dynamics for an energetic stochastic approach~\cite{Sekimoto1998,Sekimoto2010,Seifert2008} to describe the heat exchanged between the nanoparticle and the thermal bath due to a feasible nonequilibrium initial condition. We point out that the nonequilibrium protocol investigated in this work is strongly inspired in previous experimental setups~\cite{Gieseler2012,Blickle2011}. The goal is to work out an illustrative  example of a complete thermal relaxation process of a nanosystem initially prepared in a nonequilibrium state toward its equilibrium, and to point out exact analytical results that should be useful to experimental works~\cite{Gieseler2012,Blickle2011}. By following this scheme, we are able to exactly solve the probability distribution function of nonequilibrium heat exchange between the nanoparticle and the reservoir.

The rest of this paper is outlined as follows: In Sec.~\ref{derivationsec} we specify the stochastic differential equation to our particle energy and present an exact analytical solution to the time dependent probability density function. The result is then used to describe a nonequilibrium heat transference as a function of time and to explicitly verify the XFT. Section~\ref{CLTsec} verifies the physical limit of a large number $N$ of noninteracting trapped nanoparticles and compare our results to the Central Limit Theorem predictions. Direct numerical simulations are then performed at Section~\ref{DNS} in order to validate our analytical calculations, and they are also used to check the probability of a reverse heat exchange, which would be analogous to the violation of the second law of thermodynamics at the macroscopic regime. Finally, in Sec.~\ref{conclude} we present our chief conclusions and final remarks.

\section{Stochastic thermodynamics approach}
\label{derivationsec}

\subsection{Analytical solution to the Langevin dynamics}
\label{solsec}
\begin{figure}[ht]
\includegraphics[width=3.3 in]{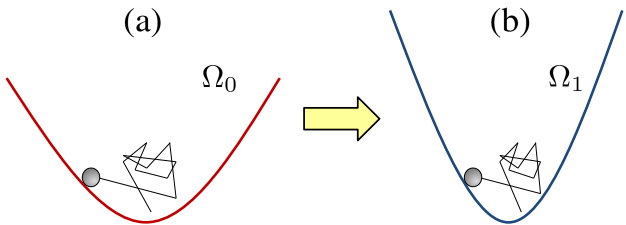}
\caption{Representative sketch of the relaxation protocol for a Brownian nanoparticle optically trapped in a harmonic potential. For $t < -\tau$ (a) the particle submitted to a laser trap of frequency $\Omega_{0}$ is at thermal equilibrium with the surrounding bath. At $t=-\tau$ the laser frequency is abruptly changed during a short time interval until it reaches the value $\Omega_{1}$ at $t=0$, and kept constant afterwards. For $t \geq 0$ (b) the particle undergoes a nonequilibrium heat exchange process with the thermal bath until it reattains the equilibrium condition.}
\label{fig1}
\end{figure}

We follow~\cite{Gieseler2012} and consider the system composed of a classical nanoparticle immersed in a low viscosity thermal reservoir of temperature $T_{2}$ and submitted to a harmonic potential produced by a laser trap, as depicted by Fig.~\ref{fig1}(a). Its Langevin dynamics can be written in the same notation as in \cite{Gieseler2012} for $\textbf{x}=\{x,y,z\}$:

\begin{equation}
    \label{Langevin}
    \ddot{\textbf{x}}(t) + \Gamma_0\dot{\textbf{x}}(t) +\Omega_0^2\textbf{x}(t) = \frac{1}{m}\textbf{F}_{fluc}(t)
\end{equation}
where the random Langevin force $\textbf{F}_{fluc}(t)$ is normally distributed with zero mean and its components satisfy $\langle F^{i}_{fluc}(t)F^{j}_{fluc}(t')\rangle = 2m\Gamma_0 k_B T_2 \delta(t-t')$ if $i=j$, and it equals zero otherwise. The $\delta(t)$ stands for the Dirac delta function and the angle brackets denote an ensemble average. The frequency $\Omega_0$ is defined as a function of the trap stiffness $k$ as $\Omega_0=\sqrt{k/m}$, $\Gamma_0$ is a friction constant, and $m$ is the particle mass. We stress that Eq.~(\ref{Langevin}) corresponds to Newton's second law applied to a vacuum-trapped levitated nanoparticle: the terms on the left-hand side represent respectively inertia, deterministic damping and the optical trap restoring force, while the term on the right-hand side accounts for the stochastic force from random molecular collisions. 

Defining the energy of the system of particles as $E(\textbf{x},\textbf{p})=1/2(m\Omega_0^2\textbf{x}^2 +\textbf{p}^2/m)$ and applying Ito's Lemma \cite{Oeksendal2003} in the highly underdamped limit $\Omega_0/\Gamma_0 \gg 1$, we follow the steps of \cite{Gieseler2012} and obtain a simple stochastic differential equation (SDE) for the total energy evolution in time:
\begin{equation}
\label{LangevinforE}
dE=-\Gamma_0 \Bigg (E-\frac{f}{2}k_B T_2 \Bigg)dt+\sqrt{2\Gamma_0 k_B T_2 E}dW_t,
\end{equation}
\newline
where $dW_t$ is the increment of the Wiener's process and we keep the $f=2d$ degrees of freedom dependency explicit, in $d$ spatial dimensions (for more details see the Appendix~\ref{Appendix2}). We point out that Eq.~(\ref{LangevinforE}) describes the energy stochastic dynamics of a nanoparticle laser-trapped in a very dilute gas (vacuum) where inertia was not neglected, in contrary to other overdamped approaches~\cite{ZonPRL2003,Zon2003,Zon2004,Blickle2006,Baiesi2006}.

Although the studied process has a linear drift, it corresponds to a diffusion problem with a multiplicative noise of the square root type and may not be considered a trivial stochastic system. Furthermore, we will show that it displays FT time-asymmetric behavior for all time, something which is usually not addressed in literature~\cite{Seifert2012,Crooks2007}. Moreover, our Langevin dynamics is in agreement with the one described by the experimental work of Ref.~\cite{Gieseler2012} in the absence of feedback cooling and for $f=2$ (see Supplementary Information of~\cite{Gieseler2012}, equation $30$). In Eq.~(\ref{LangevinforE}), the first contribution on the right-hand side is a linear growth term driving the energy with an exponential decay in time to the constant value of $\frac{f}{2}k_B T_2$, which is in agreement with the equipartition theorem from thermodynamics. However, the noisy contribution of the SDE, which is given by the term proportional to $\sqrt{T_2 E}$, prevents the energy to decay deterministically. Also, notice that the increment $dE$ is always greater than zero for $E=0$, implying that $E(t)$ must remain nonnegative. Actually, the condition $f\geq2$ ensures $E(t)$ remains positive for all time~\cite{Feller1951}. Finally, $E(t)$ is a continuous random variable and its probability density function (PDF) evolves smoothly in time, and since we are dealing with a thermal relaxation
process, we expect it to reach the equilibrium given by the known Maxwell-Boltzmann (MB) type of PDF from statistical mechanics.

In fact, in order to have analytical access to the nonequilibrium time-evolving PDF of the problem, one must solve the Fokker-Planck equation associated to the SDE~(\ref{LangevinforE}) for the transition probability $P_t(E|E_0)$ [see Eqs.~(\ref{APeq10}) and~(\ref{APeq11}) in Appendix~\ref{Appendix2}]. We should regard that, in general, $P_t(E|E_0)$ is unknown and to determine it one would have to know the Green's function of the Fokker-Planck equation for the energy. Amazingly enough, in our present situation, this is straightforward since we can readily verify that Eq.~(\ref{LangevinforE}) resembles the Cox-Ingersoll-Ross (CIR) model for interest rates \cite{CIR1985} from quantitative finance, which utilizes the general exact solution to the associated Fokker-Planck equation originally developed by Feller~\cite{Feller1951}. We point out that this analytical solution can be seen as a generalization of the one-dimensional diffusion problem~\cite{Uhlenbeck1930} with a multiplicative noise of the square root type. Although this solution
has been known for several decades, very few examples of physically motivated nonequilibrium problems have exploited it so far~\cite{Dornic2005,Evans2015}. Notably, the field of nonequilibrium thermodynamics is not applying the solution to address the fluctuation theorems where its use would enlighten experimental results in transient conditions~\cite{Gieseler2012}.

Therefore, we are able to attain the nonequilibrium time-dependent probability distribution function $P_{t}(E|E_0)$ to our SDE by simply importing the general solution from Refs.~\cite{Feller1951,CIR1985}. By performing a suitable transformation of parameters, for $t>0$ the nonequilibrium PDF is given by
\begin{eqnarray}
\label{PtE}
P_t(E|E_0)&=&c_t e^{-c_t(E+E_0 e^{-\Gamma_0 t})}\Bigg (\frac{E}{E_0 e^{-\Gamma_0 t}} \Bigg)^{q/2} \times \nonumber \\
&\times& I_q(2c_t\sqrt{E E_0 e^{-\Gamma_0 t}}),
\end{eqnarray}
for $E\geq0$ and $E_0\geq0$, where $q=f/2-1$ and
\begin{equation}
\label{constant}
c_{t}=\frac{1}{(1-e^{-\Gamma_0 t})k_B T_2}.
\end{equation}
Furthermore, $I_q(x)$ is the modified Bessel function~\cite{Arfken2012}
\begin{equation}
\label{Bessel}
I_q (x)=\sum_{m=0}^{\infty} \frac{1}{m!\Gamma(m+q+1)}\Bigg (\frac{x}{2} \Bigg )^{2m+q},
\end{equation}
and $P_{0}(E|E_0)=\delta(E-E_0)$. We point out that, in statistics, the distribution shown in (\ref{PtE}) is commonly known as the Noncentral Chisquared distribution. As expected, the equilibrium limit ($t\rightarrow\infty$) results in the MB distribution for the energy
\begin{equation}
\label{MB}
P_{\infty}(E|E_0)=\frac{\beta^{1+q}}{\Gamma(q+1)}E^{q}e^{-\beta E},
\end{equation}
for $E\geq0$, where in our case $\beta=(k_B T_2)^{-1}$, and it clearly does not depend on the initial condition $E_0$.

Now, since we possess the conditional probability of $E$ at time $t$ provided it was $E_0$ at time $0$ in Eq.~(\ref{PtE}), we are in position of applying this result to any given initial situation involving the trapped nanoparticle in a thermal reservoir. Aiming the study of a nonequilibrium heat exchange process, we adopt the protocol illustrated in Fig.~\ref{fig1}: for $t < -\tau$ (Fig.~\ref{fig1}(a)) the particle is at thermal equilibrium with a reservoir of temperature $T_{2}$ and submitted to a laser trap of frequency $\Omega_{0}$; then, during a short interval $-\tau \leq t \leq 0$ the laser frequency is abruptly changed to $\Omega_{1}$, and kept constant afterwards for $t>0$ (Fig.~\ref{fig1}(b)). As a consequence, the particle assumes a nonequilibrium MB distribution of effective temperature $T_{eff}=T_{1}$ at $t=0$, as it is demonstrated in the Appendix~\ref{Appendix1}. Therefore, for $t \geq 0$ the particle undergoes a nonequilibrium heat exchange process with the thermal bath until it reattains the equilibrium condition $T_{eff}=T_{2}$. We point out that during the whole protocol the nanoparticle is in contact with the environmental thermal bath of temperature $T_{2}$ as in Ref.~\cite{Gieseler2012}.

As a matter of fact, the system is initially prepared at $t=0$ with an effective temperature $T_1$, and then for $t\geq 0$ it relaxes in thermal contact with the reservoir with temperature $T_2$ for a time interval $t$. One may write the final energy distribution as the superposition of Eq.~(\ref{PtE}) with $\beta=\beta_2$ over the initial conditions from Eq.~(\ref{MB}) with $\beta=\beta_1$:
\begin{eqnarray}
\label{PtE2}
P_t(E)&=&\int_{0}^{\infty}P_t^{(2)}(E|E_0)P_{\infty}^{(1)}(E_0)dE_0 \nonumber \\
&=&\frac{\beta_{t}^{1+q}}{\Gamma(q+1)}E^qe^{-\beta_{t}E},
\end{eqnarray}
which is also a MB distribution for all $t>0$, where $\beta_{t}=(k_B T_t)^{-1}$ and $T_{t}$ is a time-dependent effective temperature given by
\begin{equation}
\label{betaeff}
T_{t}=T_2+(T_1-T_2)e^{-\Gamma_0 t}.
\end{equation}
The upper index $i$ in $P_t^{(i)}$ relates to the inverse temperatures $\beta_1$ and $\beta_2$ for $i=1,2$ respectively. The transition probability in Eq.~(\ref{PtE}) acts as a functional operator taking a MB equilibrium distribution at $t=0$ [Eq.~(\ref{MB})] with $\beta=\beta_1$ into another MB distribution with $\beta=\beta_2$, as expected from a statistical mechanics standpoint. However, in this specific system, the obtained nonequilibrium distribution (\ref{PtE2}) for $t>0$ is also MB, with an effective temperature $T_{t}$ that decays from $T_1$ to $T_2$. The exponential relaxation of the effective temperature in Eq.~(\ref{betaeff}) in an outcome consistent with the Newton's Law of cooling~\cite{Nath2007}.

Although the distribution given in Eq.~(\ref{PtE2}) is simple, calculating the distribution for the heat exchanged $P_t(\Delta E=E-E_0)$ requires additional steps and it consists on the main result of this paper. One starts writing the heat exchange distribution in terms of the conditional probability in Eq.~(\ref{PtE}) with $\beta=\beta_2$ and integrates over the initial conditions $E_0$ from Eq.~(\ref{MB}) with $\beta=\beta_1$ as follows:
\begin{eqnarray}
\label{Bayes}
P_{t}(\Delta E)&=&\int_{\frac{|\Delta E|}{2}}^\infty P_t^{(2)} \Bigg (E_0+\frac{\Delta E}{2}|E_0 - \frac{\Delta E}{2} \Bigg ) \times \nonumber \\
&\times& P_{\infty}^{(1)} \Bigg (E_0-\frac{\Delta E}{2} \Bigg ) dE_0.
\end{eqnarray}
The integration variable in Eq.~(\ref{Bayes}) was rewritten as $E_0-\Delta E/2$ so that final and initial energies in the superposition integral above are $E_0 \pm \Delta E /2$. This change of variable will explore a property from Eqs.~(\ref{PtE}) and (\ref{MB}) that will make XFT easily seen. Moreover, the lower bound in the integral comes from the condition $E\geq0$ in Eq.~(\ref{PtE}), after combining the inequalities $E_0+\Delta E/2 \geq0$ and $E_0-\Delta E/2\geq0$, which leads to $E_0\geq|\Delta E|/2$.

Actually, the same method may be applied to find the time-dependent nonequilibrium PDF for the energy of any free Langevin dynamics initially prepared in a known steady state, such as the experimental setup of Ref.~\cite{Gieseler2012}, provided that the integral in Eq.~(\ref{Bayes}) is solved. In the next section, Eq.~({\ref{Bayes}}) will be explicitly calculated for a system of $N$ noninteracting trapped nanoparticles submitted to our relaxation protocol (Fig.~\ref{fig1}).

\subsection{Heat exchange between the nanosystem and the reservoir and verification of the fluctuation theorem}
\label{XFTsec}
The dynamical relaxation protocol described in the previous section considers that the nanosystem is taken away from equilibrium with the reservoir of temperature $T_2$. Furthermore, the initial PDF $P^{(1)}$ at $t=0$ is the Maxwell-Boltzmann distribution (\ref{MB}) with $T=T_1$. For $t \geq 0$ the nanoparticle exchanges heat with the reservoir $T_2$ towards equilibrium. Similarly, the final nonequilibrium PDF $P_t^{(2)}$ is the time dependent PDF (\ref{PtE}) with $T=T_2$. By replacing both PDFs in Eq.~(\ref{Bayes}) and integrating over all possible initial $E_0$ one obtains:
\begin{equation}
\label{Bayess2}
P_t(\Delta E)=e^{\frac{(\beta_1-\beta_2)\Delta E}{2}}f_t(|\Delta E|),
\end{equation}
where $f_t(|\Delta E|)$ is further specified in Eq.~(\ref{Bayes2}). For consistency with Eq.~(\ref{FT1}), we define the heat flowing from the nanosystem to the reservoir by $Q=-\Delta E$. In point of fact, $Q$ is the net heat exchanged between the nanosystem $1$ and the reservoir $2$. Moreover, by dividing $P_t(-\Delta E)/P_t(+\Delta E)$, the even function $f_t(|\Delta E|)$ is canceled out and the XFT (\ref{FT1}) becomes explicitly verified as
\begin{equation}
\label{FT2}
\frac{P_t(-\Delta E)}{P_t(+\Delta E)} = e^{(\beta_2-\beta_1) \Delta E},
\end{equation}
as we wanted to demonstrate. Although $f_t(|\Delta E|)$ is system dependent, Eq.~(\ref{Bayess2}) is a general formula and it was obtained in \cite{Jar2004}. For the Langevin dynamics in the heat exchange protocol considered in this paper, the function $f_t(|\Delta E|)$ can be written from Eq.~(\ref{Bayes}) as
\begin{equation}
\label{Bayes2}
f_t(|\Delta E|)= \frac{\beta_1^{1+q}c_t^{1+q}}{\Gamma(q+1)}\sum_{m=0}^{\infty}\frac{c_t^{2m}e^{-m\Gamma_0 t}G_m(\Delta E)}{m!\theta_t^{2(m+q)+1}},
\end{equation}
where
\begin{eqnarray}
\label{Bayes2Complemento}
G_m(\Delta E)&=&\frac{\theta_t^{2(m+q)+1}}{\Gamma(m+q+1)}\int_{\frac{|\Delta E|}{2}}^{\infty} \Bigg ( E_0^2-\frac{\Delta E^2}{4} \Bigg )^{m+q} \times \nonumber \\
&\times& e^{-\theta_t E_0}dE_0,
\end{eqnarray}
and $\theta_t=\beta_1+\beta_2 \coth(\Gamma_0 t/2)$. The integral above may be written in terms of the modified Bessel function of the second kind \cite{Gradshteyn}:
\begin{eqnarray}
\label{Bayes2ComplementoB}
G_m(\Delta E)=\frac{1}{\sqrt{\pi}}(\theta_t|\Delta E|)^{m+q+1/2}K_{m+q+1/2}\Bigg(\frac{\theta_t|\Delta E|}{2} \Bigg). \nonumber \\
\end{eqnarray}
Inserting Eq.~(\ref{Bayes2ComplementoB}) in Eq.~(\ref{Bayes2}) and applying the multiplication theorem for Bessel functions \cite{Abramowitz} results in:
\begin{eqnarray}
\label{finalF}
f_t(|\Delta E|) &=& \frac{(\beta_1c_t)^{1+q}}{\sqrt{\pi}\Gamma(q+1)}\Bigg(\frac{|\Delta E|}{\ell_t\theta_t}\Bigg)^{q+1/2} \times \nonumber \\
&\times& K_{q+1/2}\Bigg(\ell_t\theta_t\frac{|\Delta E|}{2}\Bigg),
\end{eqnarray}
where $\ell_t$ is defined as
\begin{equation}
\label{ELL}
\ell_t=\sqrt{1-\frac{4c_t^2e^{-\Gamma_0t}}{\theta_t^2}}
\end{equation}

The equilibrium limit may be carried out explicitly by replacing $c_\infty=\beta_2$ and $\theta_\infty=\beta_1+\beta_2$ in Eq.~(\ref{finalF}):
\begin{eqnarray}
\label{finalPDFequilibrium}
f_\infty(|\Delta E|)&=&\frac{\beta_1^{1+q}\beta_2^{1+q}|\Delta E|^{q+1/2}}{\sqrt{\pi}(\beta_1+\beta_2)^{q+1/2}\Gamma(q+1)}\times \nonumber \\
&\times&K_{q+1/2}(\frac{|\Delta E|}{2}(\beta_1+\beta_2)).\nonumber \\
\end{eqnarray}

\begin{figure}[ht]
\includegraphics[width=3.0 in]{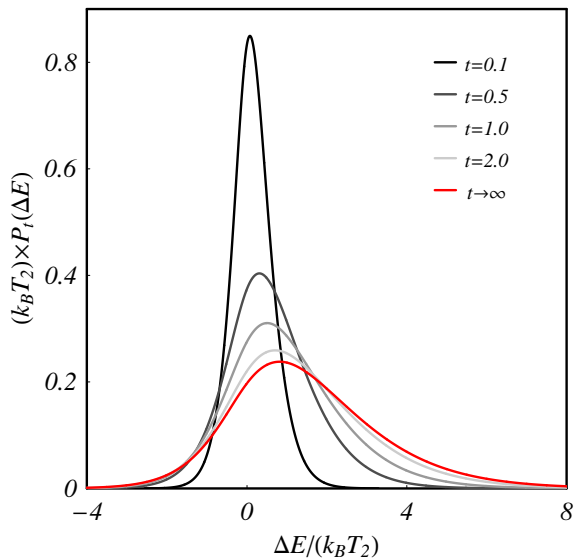}
\caption{(Color online) Relaxation dynamics of the probability distribution function (PDF) of a single nanoparticle as a function of the energy for a heating protocol with $\alpha=0.5$ and $t=0.1,0.5,1.0,2.0$, where $t$ is given in units of $1/\Gamma_{0}$. The equilibrium PDF for $t\rightarrow\infty$ is also displayed for comparison.}
\label{fig2}
\end{figure}

\begin{figure}[ht]
\includegraphics[width=3.0 in]{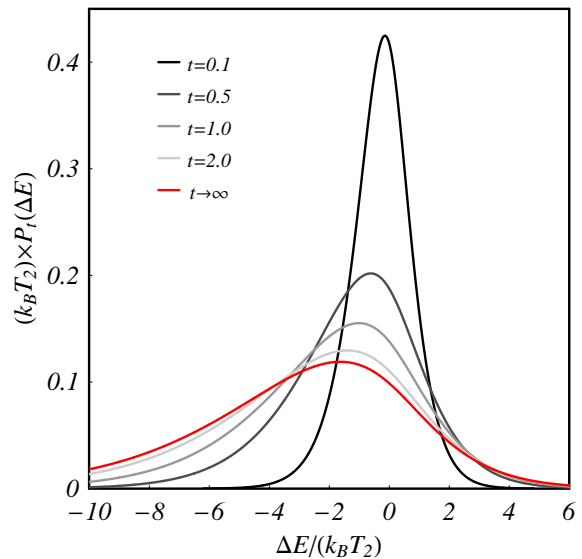}
\caption{(Color online) Relaxation dynamics of the probability distribution function (PDF) of a single nanoparticle as a function of the energy for a cooling protocol with $\alpha=2.0$ and $t=0.1,0.5,1.0,2.0$, where $t$ is given in units of $1/\Gamma_{0}$. The equilibrium PDF for $t\rightarrow\infty$ is also displayed for comparison.}
\label{fig3}
\end{figure}

Figures~\ref{fig2} and~\ref{fig3} depict the nonequilibrium PDF in Eq.~(\ref{Bayess2}) with $f_t$ given by Eq.~(\ref{finalF}), for a trapped nanoparticle in $d=3$ dimensions. The important dimensionless parameter here is the ratio $\alpha=T_1/T_2$ between the initial and final bath temperatures, which is taken as $\alpha=0.5$ in Fig.~\ref{fig2} and $\alpha=2.0$ in Fig.~\ref{fig3}. In addition, in both Figures and for the rest of our results, time is given in units of $1/\Gamma_{0}$, energy in units of $k_{B} T_{2}$, and probability densities in units of $(k_{B} T_{2})^{-1}$.
Different time instants $t=0.1,0.5,1.0,2.0$ of the PDF are compared for heating ($\alpha=0.5$) and cooling ($\alpha=2.0$) exchange protocols. The equilibrium PDFs with $f_t$ obtained in Eq.~(\ref{finalPDFequilibrium}) are also displayed for comparison. In both situations, notice the PDFs are similarly sharp for a small time interval, with relevant possibility of positive and negative energy fluctuations. As time goes by and the particle evolves to reattain thermal equilibrium with the reservoir, the PDFs start to show notably different behavior. When heating ($\alpha=0.5$), positive energy fluctuations are favored. Alternatively, negative energy fluctuations are more likely to happen when cooling ($\alpha=2.0$), as expected from a thermodynamics standpoint. It is also clear that thermal equilibrium is nearly reached after a time interval of few units of $\Gamma_0^{-1}$. In all time dependent cases and also in equilibrium, XFT holds and makes the PDFs asymmetrical with respect to $\Delta E=0$.

We close this section by emphasizing the main differences between the conditions of applications between our heat exchange protocol displayed in Fig.~\ref{fig1} and the one proposed in the original XFT work~\cite{Jar2004}. The protocol observed in Ref.~\cite{Jar2004} states that: two systems separately prepared in equilibrium with temperatures $T_{1}$ and $T_{2}$ are placed in thermal contact for a defined time $t$ and then separated again, and that the heat $Q$ exchanged in this process obeys Eq.~\ref{FT1}. Therefore, that paper was concerned with universal aspects of classic and quantum heat transfer that do not depend on the origin of the interaction between the bodies. In our current work, to exemplify the heat fluctuation in full detail through a microscopic stochastic dynamics, there is a need for a probe system to perform the heat transfer. In this case, our system $1$ is formed by a Langevin nanoparticle, and system $2$ is the thermal bath $T_{2}$. By following our relaxation protocol (Fig.~\ref{fig1}), the nanoparticles are prepared in an initial state $T_{eff}=T_{1}$ and exchange heat with the reservoir until equilibrium is reached. Naturally the assumptions of the XFT include the case considered here as a particular case, as we have verified in Eq.~\ref{FT2}. Moreover, it is important to notice that our heat exchange setup is inspired in actual feasible feedback cooling experiments of optically trapped nanoparticles~\cite{Gieseler2012}, where the thermal relaxation process of a trapped nanoparticle in contact with a thermal bath can be observed.

\section{Comparison with the central limit theorem}
\label{CLTsec}

In order to broaden our study, in this section, we investigate how our exact results for an arbitrary number of Brownian particles can be compared to the Central Limit Theorem (CLT) prediction. The CLT corresponds to the commonly used Gaussian approximation for PDFs describing macroscopic systems at thermal equilibrium, and deviations from it should be expected for small systems at nonequilibrium conditions. As it can be further seen, the XFT statement is not fulfilled by the Gaussian approximation for a thermal relaxation process, and one should use our analytical result~(\ref{Bayess2}) in order to describe properly the nonequilibrium features of small heat exchanging systems.

\begin{figure}
\includegraphics[width=3.0 in]{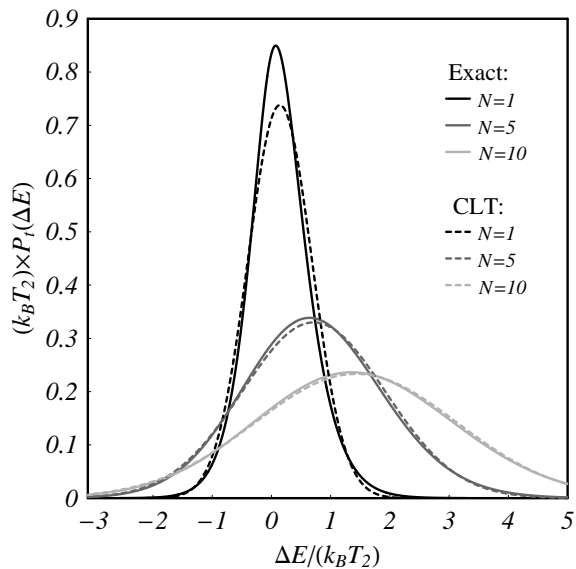}
\caption{Probability distribution of the exchanged energy $\Delta E$ for different numbers of nanoparticles $N=1,5,10$. The continuous lines correspond to the exact analytical result given by Eq.~(\ref{Bayess2}) and the dashed lines correspond to the Central Limit Theorem (Gaussian approximation) given by Eq.~(\ref{Normal}). All the curves are calculated at a fixed time $t=0.2$ (in units of $1/\Gamma_{0}$), $\alpha=0.5$ and $f=6$.}
\label{fig4}
\end{figure}

The $N$ independent trapped nanoparticles case can be obtained simply by replacing $f\rightarrow Nf$ in Eqs.~(\ref{LangevinforE})-(\ref{finalF}) due to statistical independence of the particles energies. For a large number of particles, $N\gg1$, the CLT gives an approximation for Eq.~(\ref{Bayess2}) in terms of a Normal distribution, $P_t^{CLT}(\Delta E)$, given by
\begin{equation}
\label{Normal}
P_t^{CLT}(\Delta E)=\frac{1}{\sigma_t\sqrt{2\pi}}e^{\frac{-(\Delta E-\mu_t)^2}{2\sigma_t^2}},
\end{equation}
with mean, $\mu_t=\langle\Delta E\rangle$, and variance, $\sigma_t^2=\langle\Delta E^2\rangle-\langle\Delta E\rangle^2$, obtained directly from Eqs.~(\ref{LangevinforE}) and (\ref{MB}) after some manipulation:
\begin{eqnarray}
\label{MeanE}
\mu_t&=&\frac{Nfk_B}{2}(T_2-T_1)(1-e^{-\Gamma_0 t}), \\
\sigma_t^2&=&\frac{Nfk_B^2}{2}[(T_2+(T_1-T_2)e^{-\Gamma_0t})^2 + \nonumber \\
&+&T_1^2(1-2e^{-\Gamma_0t})].
\end{eqnarray}
Notice that for $t=0$, one has $\mu_0=0$ and $\sigma_0=0$ consistent with $P_0=\delta(\Delta E)$. The expressions are also consistent with $t\rightarrow\infty$ where $\Delta E=E_2-E_1$ may be seen as the difference of two independent random variables with MB distributions (\ref{MB}) at different temperatures. Trying to access XFT for this Gaussian approximation leads to a different outcome:
\begin{equation}
\label{XFTNormal}
\frac{P_t^{CLT}(-\Delta E)}{P_t^{CLT}(+\Delta E)}=\exp \Bigg ({\frac{-2\mu_t}{\sigma_t^2}\Delta E} \Bigg),
\end{equation}
which clearly does not satisfy the condition for nonequilibrium heat exchange in Eq.~(\ref{FT2}), even though the ratio $2\mu_t / \sigma_t^2$ does not depend on $N$. This was already expected since Eq.~(\ref{XFTNormal}) deals with values of (\ref{Normal}) far from the mean of the distribution, where the CLT approximation is known to perform poorly. In order to illustrate this fact, in Fig.~\ref{fig4} we plot theoretical PDFs calculated from our exact non-Gaussian result~(\ref{Bayess2}) in continuous curves, and the CLT approximation~(\ref{Normal}) in dashed lines, for different values of particle numbers $N=1,5,10$. Notice that the CLT misses the energy fluctuation PDF for $N=1$ by a large amount, as expected. Increasing the number of particles makes CLT visually consistent with the theoretical energy fluctuation PDFs at $N=5$ and $N=10$, however the approximation does not obey XFT as observed in Eq.~(\ref{XFTNormal}). In the large fluctuation limit, the theoretical PDFs have an exponentially damping factor
in $\Delta E$, but the CLT approximations decay even faster as it is implied by the normal distribution (\ref{Normal}).

\section{Monte Carlo Simulation}
\label{DNS}

\begin{figure}
\includegraphics[width=3.0 in]{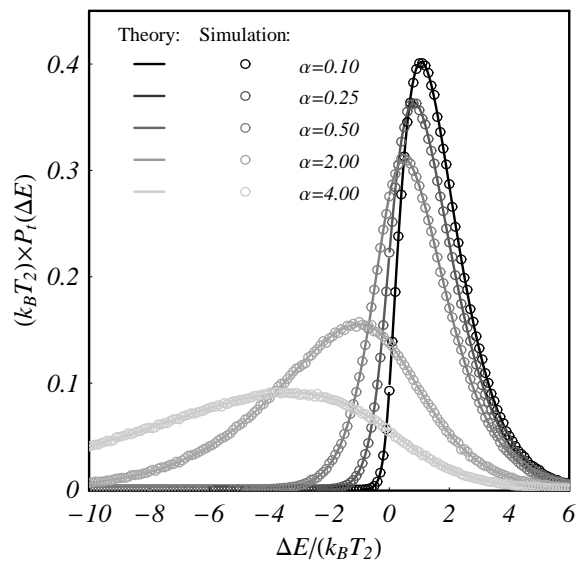}
\caption{Probability density function of $\Delta E$ evaluated from theoretical predictions (continuous lines) and Monte Carlo simulations (circles) for different values of $\alpha=0.10,0.25,0.50,2.00$ and $4.00$. All the curves are calculated at a fixed time $t=1.0$ (in units of $1/\Gamma_{0}$), $N=1$ and $f=6$.}
\label{fig5}
\end{figure}
 
Aiming the confirmation of the theoretical prediction given by Eq.~(\ref{Bayess2}), numerical Monte Carlo simulations are performed. We stress that these simulations are carried out only for corroborating the exact analytical results, and therefore they are used here only as a validation of the exact solution and for illustrative purposes. The algorithm randomly generates an initial system with energy $E_0$ from the Maxwell-Boltzmann distribution (\ref{MB}) with temperature $T_1$. Then, it starts the iterations of the Langevin stochastic dynamics (\ref{LangevinforE}) with time increment $dt$, initial energy $E_0$ and temperature $T_2$. After several time steps, the final energy $E_F$ is computed. Finally, the results of $n=10^6$ initializations are aggregated to assess the empirical PDF of $\Delta E=E_F-E_0$.

Since the energy is nonnegative, we have used Milstein's integration method \cite{Milstein1975} for the stochastic differential equation (\ref{LangevinforE}) to avoid issues around $E=0$. In this case, the discrete dynamics of the energy increments, $E_{s+1}=E_s+dE_s$, are given by
\begin{eqnarray}
\label{Milstein}
dE_s&=&-\Gamma_0(E_s-\frac{f}{2}N k_B T_2)dt+\sqrt{2bE_s}dW_s + \nonumber \\
&+&\frac{b}{2}(dW_s^2-dt),
\end{eqnarray}
where $s$ represents the discrete time instants, and we have considered $f=6$ and $b=\Gamma_0 k_B T_2$. The simulation was performed with time steps $dt=10^{-2}\Gamma_0^{-1}$ and 
$dW_s$ being obtained from the Normal Distribution with null mean and variance $dt^2$. Notice that the last term in Eq.~(\ref{Milstein}) gives the Milstein's higher order correction to the Euler$-$Maruyama method and it is simple for the SDE of Eq.~(\ref{LangevinforE}). After $t/dt$ iterations of Eq.~(\ref{Milstein}) for each of the $n=10^6$ initializations, the final energy $E_F$ is calculated and the histogram of $\Delta E = E_F - E_0$ is evaluated.

A different approach would be to integrate the velocity dynamics in Eq.~(\ref{Langevin}) for each degree of freedom, but it would be less efficient than ours. A more computationally efficient simulation of the energy PDF based on \cite{Dornic2005} would draw $E_F$ from the distribution (\ref{PtE}) directly from a mixture of a Gamma and a Poisson number generators for any instant of time. Together with the usual draw of $E_0$ from a MB distribution, which is also a Gamma distribution, it would lead to $\Delta E$ and avoid the $t/dt$ time iterations. However, our integration approach has the advantage of generating single trajectories which would permit time dependent parameters naturally, such as external time dependent potentials~\cite{Dieterich2015}, to be explored in further simulations.

Figure~\ref{fig5} shows a comparison between the nonequilibrium PDF calculated analytically from Eqs.~(\ref{Bayess2}) and (\ref{finalF}), represented by continuous curves, and the Monte Carlo simulation of the stochastic dynamics in (\ref{Milstein}), illustrated by circles, for several values of $\alpha$. By analyzing both results, we observe the theory is remarkably validated by the simulations in both cooling ($\alpha>1$) and heating ($\alpha<1$) dynamic protocols, where the energy fluctuations are biased towards negative and positive values, respectively, depicting two notably distinct types of asymmetries where XFT holds.

\subsection{Probability of reverse heat exchange}
\label{Violationsection}

In this section, we take a closer look at the energy flow direction in the heat exchange process. For the case of $T_2>T_1$, or $\alpha<1$, one should expect $\Delta E>0$ from a thermodynamical point of view. This would be analogous to a positive entropy production $(\beta_1 - \beta_2)\Delta E > 0$ of the net heat exchanged between the nanosystem $1$ and the reservoir $2$. However, invoking a more general entropy expression produced by our Langevin nanoparticle can be tricky~\cite{Seifert2005} and it is not even necessary to our calculations. As a matter of fact, in this stochastic approach, since the energy fluctuation is seen as a random variable, there is a chance of a reverse heat flow, e. g., a flow from cold particles to a hotter bath. At the level of macroscopic thermodynamics (and in the absence of external work), the passage of heat from a colder to a hotter body constitutes a violation of the second law~\cite{Jar2004}. From our analytical solution to the problem, we can derive the probability of observing such a \textquotedblleft violation\textquotedblright. Strictly speaking, in this framework, the second law is commonly restated as an ensemble averaged law and it is satisfied even for $N=1$. But as the random variable $\Delta E$ may take any possible value, the probability for a negative entropy variation may be calculated from theory. First, consider the cumulative density function (CDF) of (\ref{PtE}) evaluated at $E_0$:
\begin{equation}
\label{Violation1}
P_t(\Delta E<0|E_0)=\int_{0}^{E_0}P_t(E|E_0)dE.
\end{equation}
\begin{figure}
\includegraphics[width=3.0 in]{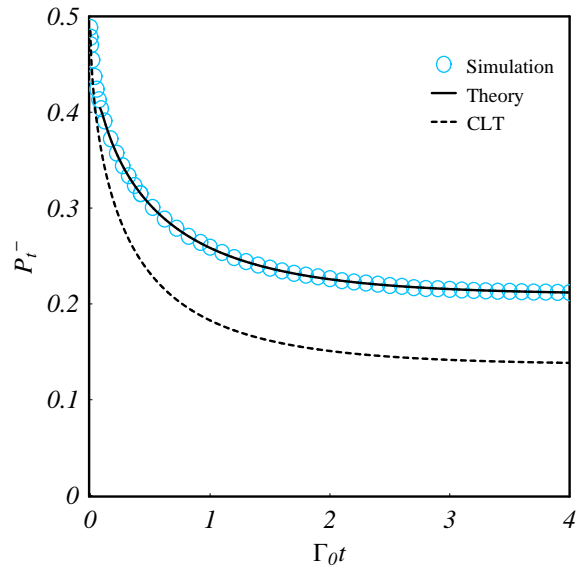}
\caption{Probability of reverse heat flow as a function of the dimensionless time $\Gamma_0 t$, for $N=1$, $f=6$ and $\alpha=0.5$. The plot depicts the probability of a single nanoparticle to lose energy in a heating process. Our theoretical calculation given by Eq.~(\ref{Violation3}) is represented by the continuous line, whereas Monte Carlo simulations are shown by circles and the CLT prediction (Gaussian approximation) is exhibited  by the dashed curve.}
\label{fig6}
\end{figure}
Using the CDF of a Noncentral Chi-squared distribution, one obtains:
\begin{eqnarray}
\label{Violation2}
P_t(\Delta E<0|E_0)&=&e^{-c_tE_0(1+e^{-\Gamma_0t})} \times \nonumber \\
&\times& \sum_{i,j=0}^{\infty}\frac{e^{-j\Gamma_0t}(c_tE_0)^{i+2j+\frac{Nf}{2}}}{j!\Gamma(Nf+j+i+1)}.  \nonumber \\
\end{eqnarray}
Averaging over all possible initial values $E_0$ leads to the probability of a reverse heat flow as a function of time
\begin{equation}
\label{Violationdummy}
P_t^{-}=\int_{0}^{\infty} P_t(\Delta E<0|E_0)P_{\infty}(E_0)dE_0,
\end{equation}
where we defined $P_{t}^{-}\equiv P_t(\Delta E < 0)$. Combining (\ref{Violation2}) and (\ref{MB}) in the integral above leads to
\begin{eqnarray}
\label{Violation3}
P_t^{-}&=&\frac{\beta_1^{\frac{Nf}{2}}c_t^{\frac{Nf}{2}}}{\Gamma(\frac{Nf}{2})\theta_t^{Nf}} \times \nonumber \\
&\times& \sum_{i,j=0}^{\infty}\frac{e^{-j\Gamma_0 t} \Gamma(Nf+2j+i)}{j!\Gamma(\frac{Nf}{2}+j+i+1)} \Bigg (\frac{c_t}{\theta_t} \Bigg)^{i+2j}. \nonumber \\
\end{eqnarray}

Figure~\ref{fig6} shows the theoretical probability of reverse heat flow calculated in Eq.~(\ref{Violation3}) as a function of time, represented by the continuous curve, and the equivalent Monte Carlo simulation generated from Eq.~(\ref{Milstein}), given in circles, for $N=1$, $f=6$ and $\alpha=0.5$. As the considered value of $\alpha$ implies $\beta_1-\beta_2>0$, the expected outcome of the process would be $\langle\Delta E\rangle>0$ for the averaged version of the Second Law to hold. But the energy fluctuation is not deterministically positive and there is a chance of a negative fluctuation $\Delta E<0$ to take place. In fact, the probability of negative energy fluctuation rapidly decays from its initial value of $0.5$ at $t=0$ to a nonzero stationary equilibrium probability. For a complete comparison, in Fig.\ref{fig6} we also exhibit the corresponding CLT approximation extracted from Eq.~(\ref{Normal}) and depicted by the dashed line. As it should be expected, the CLT approximation significantly
underestimates the probability of violation of the second law for the $N=1$ particle case.

Furthermore, one may verify that the asymptotic limit $t\rightarrow\infty$ turns Eq.~(\ref{Violation3}) into a simpler form:
\begin{eqnarray}
\label{ViolationEquilibrium}
P_\infty^{-}=\frac{\alpha^{\frac{Nf}{2}}}{\Gamma(\frac{Nf}{2})(1+\alpha)^{Nf}}\sum_{i=0}^{\infty}\frac{\Gamma(Nf+i)}{\Gamma(\frac{Nf}{2}+i+1)}\Bigg(\frac{\alpha}{\alpha+1}\Bigg)^{i}, \nonumber \\
\end{eqnarray}
which represents the constant value of the probability of reverse heat exchange assumes for large $t$, as it can be seen in Fig.~\ref{fig6}. Moreover, as the number of particles $N$ is increased, this asymptotic value decreases and approaches to zero. This means that, as the system becomes larger, the probability of violating the second law for long times becomes smaller, and one should expect it to vanish in the thermodynamic limit $N\rightarrow\infty$. By using Stirling's approximation in the gamma functions above, it is possible to obtain an interesting relation for the system in the thermodynamic limit:
\begin{equation}
\label{ViolationLimit}
\lim_{N\rightarrow\infty} -\frac{1}{Nf}\ln P_{\infty}(\Delta E<0)=\ln\Bigg(\frac{1+\alpha}{2\sqrt{\alpha}}\Bigg)\equiv \ln A_\alpha,
\end{equation}
where $A_\alpha=(1+\alpha)/(2\sqrt{\alpha})$ is responsible for an exponential multiplicative damping factor of the type $P_{\infty}(\Delta E<0)\propto A_{\alpha}^{-Nf}$. For very small systems ($N\approx1$) a negative energy fluctuation will take place more often. However, for larger systems ($N\gg1$) the exponential decay $A_{\alpha}^{-Nf}$ will eliminate the possibility of negative energy fluctuations. Also, notice that for the situation where the  temperatures of both reservoirs are equal $\alpha=1$, the expression above vanishes since $\ln A_1=0$, so negative and positive fluctuations are equally likely to happen in the thermodynamic limit. However, in the heating protocol $0<\alpha<1$, as reverse heat exchange becomes less likely, notice that $(1-\sqrt{\alpha})^2>0$ immediately implies $A_\alpha>1$. Therefore, in the heating protocol, a negative energy fluctuation is extremely rare in the thermodynamic limit, driven by the exponential penalty $A_{\alpha}^{-Nf}$ with $A_\alpha>1$, reproducing a physical outcome for the arrow of time in large systems.

\section{Summary and conclusions}
\label{conclude}

We have obtained an exact analytic solution to the Langevin energy dynamics associated to a nanoparticle in a low friction thermal bath and trapped by a harmonic potential, within an energetic stochastic scope. The system we chose to study was strongly motivated by previous experimental setups of Brownian particles laser-trapped in a thermal bath~\cite{Gieseler2012,Blickle2011}. As a matter of fact, our analytical results could be experimentally verified by combining the setup of Ref.~\cite{Gieseler2012} with the laser protocol we proposed inspired by Ref.~\cite{Blickle2011}. We stress that, usually in this class of problems, the time-dependent probability distributions for stochastic trajectories are only assessed via experimental or numerical estimations, as in Refs.~\cite{Crooks2007,Gieseler2012}. Actually, these previous works were able to display analytical results only for asymptotic probability distribution functions which characterize steady or equilibrium states. In contrary to this fact, we were able to attain the nonequilibrium time-dependent probability distribution function $P_{t}(E|E_0)$ to our SDE by importing the general solution from Refs.~\cite{Feller1951,CIR1985}. To the best of our knowledge this is the first time this SDE solution has been successfully applied to describe a physical system in a nonequilibrium thermodynamics context, and this feature of our analytical analysis sheds a new light at the general discussion on Brownian motion problems~\cite{Lee2013}.

Moreover, the general solution of the energetic stochastic dynamics was utilized for describing a nonequilibrium dynamic relaxation protocol: by abruptly switching the trap stiffness of our setup at $t=0$, our Langevin nanoparticle behaves as if it had an effective temperature $T_{1}$ which differs from the thermal bath $T_{2}$. In this way, we were able to analyze the relaxation process of the nanoparticle towards equilibrium for a nonequilibrium cooling or heating process. We have also explicitly verified that the heat exchange fluctuation theorem (XFT)~\cite{Jar2004} is indeed satisfied for our protocol, which makes our system a good textbook example of a FT application.

Furthermore, we show our analytical results are consistent with the limits of large $t$, which is represented by the equilibrium Maxwell-Boltzmann distribution, and of a large number $N$ of independent particles, which correspond to the CLT predictions. Then, direct numerical Monte Carlo simulations were performed in order to confirm the validation of our time-dependent exact solutions, and an excellent agreement between them was observed. Finally, we have computed the probability of reverse heat flow of our stochastic trajectories, which would be analogous to the violation of the second law of thermodynamics at the macroscopic regime. This violation probability becomes relevant for small systems and short time scales, as it was experimentally verified in Ref.~\cite{Evans2002}.

Finally, we expect that our analytical results may be useful to nonequilibrium stochastic systems such as colloidal particles in optical traps~\cite{Gomez2011,Mestres2014}, thermal ratchets~\cite{Prost1999}, nanomachines~\cite{Dieterich2015} and molecular motors~\cite{Seifert2012}.

\appendix

\section{Stochastic Differential Equation for the Energy}
\label{Appendix2}
In this section, the SDE for the energy (\ref{LangevinforE}) is deduced. The connection between the SDE and the underlying Fokker-Planck equation is also discussed. 

First, consider a simplified system composed of a particle in $1$ dimension, with position $x$ and momentum $p=m\dot{x}$, as in equation (\ref{Langevin}). The energy of the system is defined as
\begin{equation}
\label{APeq01}
E(x,p)=\frac{1}{2}(m\Omega_0^2 x^2+\frac{p^2}{m}).
\end{equation}
Because both position and momentum are stochastic quantities, the energy defined above is also stochastic. Its random value depends on the particle's random position and momentum simultaneously. However, one may use the equation of motion (\ref{Langevin}) to derive a equation for the energy evolution in time. In a deterministic hamiltonian system, the energy evolution is straightforward, $dE/dt=0$. But the dissipation and the random nature of the Langevin problem will allow the energy to fluctuate in time. In order to deal with differentials of stochastic quantities, one should use a framework of stochastic calculus. In this case, the tool to be used is the Ito's lemma \cite{Oeksendal2003} for the energy $E(x,p)$:
\begin{equation}
\label{APeq02}
dE=\frac{\partial E}{\partial x}dx+\frac{1}{2}\frac{\partial^2E}{\partial x^2}(dx)^2+\frac{\partial E}{\partial p}dp+\frac{1}{2}\frac{\partial^2E}{\partial p^2}(dp)^2.
\end{equation}
Notice that the equation above resembles a Taylor's expansion of the differential $dE$ in second order of $dx$ and $dp$. As both $x$ and $p$ are stochastic variables, they are not differentiable in time. Therefore, the terms in $(dx)^2$ and $(dp)^2$ may contain relevant information. Following the steps of \cite{Gieseler2012}, replacing (\ref{APeq01}) in (\ref{APeq02}) leads to

\begin{equation}
\label{APeq03}
dE=m\Omega_0^2xdx+\frac{m\Omega_0^2}{2}(dx)^2+\frac{p}{m}dp+\frac{1}{2m}(dp)^2.
\end{equation}
Now we need to use the Langevin equation (\ref{Langevin}), together with $dx=(p/m)dt$, to find a equation for to find an equation for $dp$. In this case, one obtains the stochastic equation for the momentum evolution:

\begin{equation}
\label{APeq04}
dp=(-m\Omega_0^2x-\Gamma_0p)dt+\sqrt{2m\Gamma_0k_BT_2}dW.
\end{equation}

From the definition $dx=(p/m)dt$, one gets $(dx)^2=0$ in order $dt$. To find the term in order $(dp)^2$ in (\ref{APeq03}), one should square (\ref{APeq04}) and use the identity $dW^2=dt$. After collecting the terms in order $dt$, it results in

\begin{equation}
\label{APeq05}
(dp)^2=2m\Gamma_0k_BT_2dt.
\end{equation}

Upon replacing (\ref{APeq04}) and (\ref{APeq05}) on (\ref{APeq03}), one finds the following stochastic differential equation (SDE) for the energy:

\begin{equation}
\label{APeq06}
dE=-\Gamma_0(\frac{p^2}{m}-k_BT_2)dt+\sqrt{\frac{p^2}{m}2\Gamma_0k_BT_2}dW.
\end{equation}
The dependency on $p$ above makes the stochastic equation complicated to deal with. However, one may explore the highly underdamped limit $\Gamma_0/\Omega_0\ll1$ to approximate the SDE for the energy, as presentented in \cite{Gieseler2012} with the mathematical details. Physically speaking, this limit considers the Langevin equation (\ref{Langevin}), or its equivalent (\ref{APeq04}), with very small dissipation and fluctuation terms when compared to the potential energy determined by the laser frequency. In this case, the particle behaves locally in time as a harmonic oscillator with energy $E$ over the small time interval of a laser oscillation, $\tau=2\pi\Omega_0^{-1}$. Thus it makes the virial theorem useful, $\langle p^2/m \rangle = E$, where the $\langle \rangle$ symbol is understood as the average over a time small time interval. Replacing the virial theorem approximation in (\ref{APeq06}) leads to the final form of the SDE for the energy of a single particle in one dimension:
\begin{equation}
\label{APeq07}
dE=-\Gamma_0(E-k_BT_2)dt+\sqrt{2E\Gamma_0k_BT_2}dW,
\end{equation}
which can be understood as the general SDE for the energy in the case $N=1$ and $f=2$. The introduction of a system of $N$ independent particles in $d$ dimensions is straightforward. First, notice that each dimension of each particle obey equation (\ref{APeq07}) and the total energy of the system, $E_{sys}$ , can be written as the sum of the individual energy $E_i$ of each dimension for each particle $i=1,...,Nd$:
\begin{equation}
\label{APeq08}
E_{sys}=\sum_{i=1}^{Nd}E_{i}.
\end{equation}
Taking the differential of $E_{sys}$ and using the linearity of the drift term of (\ref{APeq07}) results in
\begin{equation}
\label{APeq09a}
dE_{sys}=-\Gamma_0(E_{sys}-\frac{f}{2}Nk_BT_2)dt+\sum_{i=1}^{Nd}\sqrt{2E_i\Gamma_0k_BT_2}dW_i,
\end{equation}
where we have defined the number of degrees of freedom of each particle as $f=2d$. It is easy to see that the noise term above is a sum of gaussian random variables with zero mean. The sum is equivalent to a single gaussian increment with variance given by
\begin{equation}
\label{APeq09}
\sum_{i=1}^{Nd}\sum_{j=1}^{Nd}2\Gamma_0k_BT_2\sqrt{E_iE_j}\langle dW_idW_j\rangle=
2\Gamma_0k_BT_2E_{sys}dt,
\end{equation}
where the uncorrelated noise, $\langle dW_i dW_j \rangle = \delta_{i,j}dt$, and the definition (\ref{APeq08}) were used in the last passage. Actually, one might replace the sum of uncorrelated gaussian noises by a single gaussian noise with same mean and variance. It simplifies (\ref{APeq09}) to the final form of the SDE of the energy of a system of independent particles:
\begin{equation}
\label{APeq10}
dE=-\Gamma_0(E-\frac{f}{2}Nk_BT_2)dt+\sqrt{2E\Gamma_0k_BT_2}dW,
\end{equation}
with $E_{sys}=E$, for simplicity. Equation above is the equation (\ref{LangevinforE}) used in the present article, with $N=1$. As mentioned in the introduction, this SDE has a linear drift with a noise proportional to $\sqrt{E}$. The time dependent PDF for this equation is the solution to the following Fokker-Planck equation:

\begin{equation}
\label{APeq11}
\frac{1}{\Gamma_0}\frac{\partial{P}}{\partial{t} }=\frac{\partial}{\partial{E}}(E-\frac{f}{2}Nk_BT_2)P+\frac{\partial^2}{\partial{E^2}}(k_BT_2E)P,
\end{equation}
with $P=P(E,t)$. Fortunately, the mathematical details leading to the solution of this PDE from its Laplace transform were derived by Feller, as it is shown in Eq. ($6.2$) of Ref.~\cite{Feller1951}. In his original derivation, the author is aware that the PDE equivalent to (\ref{APeq11}) could be interpreted as the Fokker-Planck equation of a diffusion problem with a linear diffusion coefficient, $\sigma(x)^2 \propto x$. Moreover, the physically sound solution to our problem, which corresponds to the case $N f/2\geq 1$, has been utilized by the CIR model in quantitative finance, as it can be seen in Eq. ($18$) of Ref.~\cite{CIR1985}. Finally, as it is discussed in this present article, the same solution can be applied to the energy of a Langevin nanosystem.

\section{Effective Temperature Protocol}
\label{Appendix1}
Initially, the nanoparticle is optically trapped by a laser with frequency $\Omega_0$ at room temperature $T_2$, which leads to an equilibrium Maxwell-Boltzmann energy distribution with $f=2d$ degrees of freedom, where $d$ is the number of spatial dimensions. At $t=-\tau$, the laser frequency is changed linearly as $\Omega=(\Omega_1-\Omega_0)(t/\tau)+\Omega_1$ over a small time interval $\tau$, and kept constant ($\Omega=\Omega_1$) for $t \geq 0$. This fast protocol acts as an adiabatic expansion/compression and it should create the effect of a nonequilibrium MB distribution for the energy with an effective temperature $T_{eff}=T_{1}=\gamma~ T_2$ at $t=0$, for a constant $\gamma$ to be determined, as discussed below. Both frequencies are assumed to be in the highly underdamped limit $\Omega_{0,1} \gg \Gamma_0$. Also the protocol is assumed to last for a small time interval $\tau$, where $\Gamma_0^{-1}\gg\tau\gg\Omega^{-1}$. At $t=-\tau$, the protocol starts to change the energy of the particle according to the stochastic work relation \cite{Sekimoto2010}:
\begin{equation}
\label{AP2eq01}
dE=dW=\frac{\partial{U(\Omega,\textbf{x})}}{\partial{\Omega}}\frac{d\Omega}{dt}dt,
\end{equation}
where $U(\Omega,\textbf{x})$ is the potential energy and the absence of heat during the protocol follows from the low friction limit ($\Gamma_0^{-1}\gg\tau$), which is equivalent to an adiabatic transformation.
Inserting the harmonic potential, $U=m\Omega^2\textbf{x}^2/2$, and rewriting the last expression leads to
\begin{equation}
\label{AP2eq02}
dE=\Bigg(\frac{2}{\Omega}\frac{d\Omega}{dt}\Bigg)U(\Omega,\textbf{x})dt.
\end{equation}
Also from the low friction limit, the potential energy is expected to behave as $\langle U(\Omega,\textbf{x})\rangle=E/2$ due to the virial theorem, where $E$ and $\Omega$ are assumed to be approximately constant over one oscillation period $\Delta\tau=2\pi/\Omega$ \cite{Gieseler2012}. Integrating (\ref{AP2eq02}) over $\Delta\tau$ and using $d\Omega/dt=\Delta \Omega/\Delta \tau$ leads to
\begin{equation}
\label{AP2eq03}
\frac{1}{E}\frac{\Delta E}{\Delta \tau} = \frac{1}{\Omega}\frac{\Delta \Omega}{\Delta \tau},
\end{equation}
where $\Delta E = \int_{t}^{t+\Delta \tau} dE$ is an increment of $E$ over an subinterval $(t,t+\Delta\tau)$ in $(-\tau,0)$. Now using the fact that $\Delta \tau$ is short, one can solve (\ref{AP2eq03}) as a differential equation resulting in
\begin{equation}
\label{AP2eq04}
E=E_0(\Omega_1/\Omega_0)=\gamma E_0,
\end{equation}
where $E_0$ and $E$ are the stochastic energies of the particle at $t=-\tau$ and $t=0$ respectively and $\gamma=\Omega_1/\Omega_0$. Notice that $\gamma>1$ ($\gamma<1$) for $\Omega_1>\Omega_0$ ($\Omega_1<\Omega_0$) corresponds to an adiabatic heating (cooling) process during the time interval $\tau$. The protocol results in a scale transformation in the energy random variable. The scale transformation preserves the MB distribution of the energy random variable, leading to an effective temperature $T_{1}=\gamma~ T_2$. As the distribution is not in equilibrium with the room temperature $T_2$, it is expected a relaxation towards equilibrium for $t>0$. Moreover, we may readily identify that $\gamma=\alpha=T_{1}/T_{2}$. Also notice that (\ref{AP2eq04}) implies $E\Omega^{-1}$ is a constant during the protocol. By averaging the energy and using the usual definition of volume $V\propto\Omega^{-d}$ \cite{Blickle2011}, one obtains $T^{f/2}V=const.$, which is similar to the polytropic process equation of a thermodynamical adiabatic process.


\begin{thebibliography}{99}
\bibitem{Bustamante2005}C. Bustamante, J. Liphardt, and F. Ritort, Phys. Today {\bf 58}, 43 (2005). 

\bibitem{Seifert2012}U. Seifert, Rep. Prog. Phys. {\bf 75}, 126001 (2012). 

\bibitem{Velasco2011} R. M. Velasco, L. S. Garc\'{i}a-Col\'{i}n and Francisco Javier Uribe, Entropy {\bf 13}, 82 (2011). 

\bibitem{Seifert2016} U. Seifert, Phys. Rev. Lett. {\bf 116}, 020601 (2016). 

\bibitem{Evans1993}D. J. Evans, E. G. D. Cohen, and G. P. Morriss, Phys. Rev. Lett. {\bf 71}, 2401 (1993). 

\bibitem{Jar1997} C. Jarzynski, Phys. Rev. Lett. {\bf 78}, 2690 (1997). 

\bibitem{Crooks1999}G. E. Crooks, Phys. Rev. E {\bf 60}, 2721 (1999). 

\bibitem{Crooks2007}G. E. Crooks and C. Jarzynski , Phys. Rev. E {\bf 75}, 021116 (2007). 

\bibitem{Evans2015}J. Szavits-Nossan and M. R. Evans, J. Stat. Mech. {\bf 12}, P12008 (2015). 

\bibitem{Evans2002}G. M. Wang, E. M. Sevick, E. Mittag, D. J. Searles, and D. J. Evans, Phys. Rev. Lett. {\bf 89}, 050601 (2002). 

\bibitem{Carberry2004} D. M. Carberry, J. C. Reid, G. M. Wang, E. M. Sevick, Debra J. Searles, and Denis J. Evans, Phys. Rev. Lett. {\bf 92}, 140601 (2004).

\bibitem{Ritort2005}D. Collin, F. Ritort, C. Jarzynski, S. B. Smith, I. Tinoco and C. Bustamante, Nature {\bf 437}, 231 (2005). 

\bibitem{Gieseler2012}J. Gieseler, R. Quidant, C. Dellago, and L. Novotny, Nature Nanotech. {\bf 9}, 358 (2014). 

\bibitem{Blickle2011}V. Blickle, and C. Bechinger,  Nature Phys. {\bf 8}, 143 (2011). 

\bibitem{Sivak2013}D. A. Sivak, J. D. Chodera, and G. E. Crooks,  Phys. Rev. X {\bf 3}, 011007 (2013).

\bibitem{Jar2004}C. Jarzynski and D. K. W\'{o}jcik, Phys. Rev. Lett. {\bf 92}, 230602 (2004). 

\bibitem{Sekimoto1998} K. Sekimoto,  Prog. Theor. Phys. Supp. {\bf 130}, 17 (1998). 

\bibitem{Sekimoto2010} K. Sekimoto,  {\it Stochastic Energetics} (Springer, Berlin, 2010).

\bibitem{Seifert2008} U. Seifert, Eur. Phys. J. B {\bf 64}, 423 (2008). 

\bibitem{Oeksendal2003} B. Oksendal, {\it Stochastic Differential Equations: An Introduction with Applications} (Springer, Berlin, 2003).

\bibitem{Zon2003} R. van Zon, and E. G. D. Cohen, Phys. Rev. E {\bf 67}, 046102 (2003). 

\bibitem{Zon2004} R. van Zon, and E. G. D. Cohen, Phys. Rev. E {\bf 69}, 056121 (2004). 

\bibitem{ZonPRL2003} R. van Zon, and E. G. D. Cohen, Phys. Rev. Lett. {\bf 91}, 110601 (2003). 

\bibitem{Blickle2006} V. Blickle, T. Speck, L. Helden, U. Seifert, and C. Bechinger, Phys. Rev. Lett. {\bf 96}, 070603 (2006). 

\bibitem{Baiesi2006} M. Baiesi, T. Jacobs, C. Maes, and N. S. Skantzos, Phys. Rev. E {\bf 74}, 021111 (2006). 

\bibitem{Feller1951} W. Feller, Ann. Math. {\bf 54}, 173 (1951).

\bibitem{CIR1985} J. C. Cox, J. E. Ingersoll and S. A. Ross, Econometrica {\bf 53}, 385 (1985). 

\bibitem{Uhlenbeck1930} G. E. Uhlenbeck and L. S. Ornstein, Phys. Rev. {\bf 36}, 823 (1930).

\bibitem{Dornic2005} I. Dornic, H. Chate, and M. A. Munoz, Phys. Rev. Lett. {\bf 94}, 100601 (2005).

\bibitem{Arfken2012} G. B. Arfken, H. Weber, F. E. Harris, {\it Mathematical Methods for Physicists: A Comprehensive Guide} (Academic Press, 2012).

\bibitem{Nath2007} M. R. Nath, S. Sen, and G. Gangopadhyay, J. Chem. Phys. {\bf 127}, 094505 (2007). 

\bibitem{Gradshteyn} I. S. Gradshteyn, and I. M. Ryzhik, {\it Table of Integrals, Series and Products}, (New York: Academic, 2007).

\bibitem{Abramowitz} M. Abramowitz; I. A. Stegun, {\it Handbook of Mathematical Functions with Formulas, Graphs, and Mathematical Tables}, (Dover, 1972).

\bibitem{Milstein1975}  G. N. Milstein, Theory Probab. Appl. {\bf 19}(3), 557 (1975). 

\bibitem{Dieterich2015} E. Dieterich, J. Camunas-Soler, M. Ribezzi-Crivellari, U. Seifert and F. Ritort, Nature Phys. {\bf 11}, 971 (2015). 

\bibitem{Seifert2005} U. Seifert, Phys. Rev. Lett. {\bf 95}, 040602 (2005). 

\bibitem{Lee2013} J. S. Lee, C. Kwon, and H. Park, Phys. Rev. E {\bf 87}, 020104(R) (2013). 


\bibitem{Gomez2011} J. R. Gomez-Solano, A. Petrosyan, and S. Ciliberto, Phys. Rev. Lett. {\bf 106}, 200602 (2011). 

\bibitem{Mestres2014} P. Mestres, I. A. Martinez, A. Ortiz-Ambriz, R. A. Rica, and E. Roldan, Phys. Rev. E {\bf 90}, 032116 (2014). 

\bibitem{Prost1999} A. Parmeggiani, F. Julicher, A. Ajdari, and J. Prost, Phys. Rev. E {\bf 60}, 2127 (1999).


\end{thebibliography}
\end{document}